\documentclass[doublecol]{epl2} 

\usepackage{graphics,dcolumn,bm,fleqn,epic,eepic,float}
\usepackage{graphicx,epsfig}

\title{Disentangling Social and Group heterogeneities:\\ Public Goods games
  on Complex Networks} \shorttitle{Public Goods games on structured networks}

\author{J. G\'omez-Garde\~nes\inst{1,2} \and D. Vilone\inst{3} \and A. S\'anchez\inst{1,3}}
\shortauthor{J. G\'omez-Garde\~nes \etal}

\institute{                    
  \inst{1} Institute for Biocomputation and Physics of Complex Systems, Universidad de Zaragoza, E-50018 Zaragoza (Spain)\\
  \inst{2} Departmento de F\'{\i}sica de la Materia Condensada, Universidad de Zaragoza, 50009 Zaragoza (Spain)\\
  \inst{3} Departamento de Matem\'aticas, Universidad Carlos III, 28911 Legan\'es, Madrid (Spain)\\
}


\abstract{In this Letter we present a new perspective for the study of
  the Public Goods games on complex networks. The idea of our approach
  is to consider a realistic structure for the groups in which Public
  goods games are played. Instead of assuming that the social network
  of contacts self-defines a group structure with identical
  topological properties, we disentangle these two interaction
  patterns so to deal with systems having groups of definite sizes
  embedded in social networks with a tunable degree of
  heterogeneity. Surpisingly, this realistic framework, reveals that
  social heterogeneity may not foster cooperation depending on the
  game setting and the updating rule.}

\begin{document}

\maketitle

\section{Introduction}

The last few years have witnessed the success of the application of
physical techniques and concepts to social systems
\cite{castellano:2009}. One topic that has attracted a considerable
amount of attention is the evolutionary dynamics
\cite{hofbauer:1998,nowak:2006a} of social dilemmas
\cite{kollock:1998}: Situations in which an individually desirable
outcome leads to an undesirable one from a societal viewpoint. A
particularly important paradigm in this class is the tragedy of the
commons \cite{hardin:1968} or, as it is more generally known, the
Public Goods game \cite{olson:1965,ostrom:2000}. In a Public Goods
game (PGG), altruist or cooperative individuals in a group of $m$
people contribute an amount $c$ (ÒcostÓ) to the public good; selfish
people or defectors do not contribute. The total contribution is
multiplied by an enhancement factor $r < m$ and the result is equally
distributed between all $m$ members of the group.  Hence, defectors
get the same benefit of cooperators at no cost, {\em i.e.}, they
free-ride on the cooperatorsÕ effort. The dilemma then arises as
nobody has any incentive to contribute to the public good, and
therefore nobody receives any benefit. A number of hypotheses have
been put forward to explain why people might eventually contribute to
a public good, including reputation, punishment, beliefs and other
factors \cite{huberman:1994,sigmund:2010}.

One of such hypotheses is of particular relevance for our research,
namely, that contributions to a public good are enhanced by the
assortment of individuals. This implies that contributors interact
mostly with other contributors and therefore end up doing better than
free-riders. There are several roads to assortment but prominent among
them is the existence of a social network that dictates who interact
with whom. This proposal, originating \cite{nowak:1992} on work on
another paradigm, the Prisoner's Dilemma \cite{axelrod:1984}, has
given rise to an explosion of research on evolutionary game theory on
graphs \cite{szabo:2007,roca:2009a}. For the specific context of the
PGG, the issue was considered by Brandt {\em et al.}
\cite{brandt:2003}, whose numerical simulations indicated that local
interactions can foster contribution, even for values of $r$ well
below the critical value $r = m$ (above this threshold contributing is
obviously always the best option).  While this result was obtained on
an hexagonal lattice, subsequent research
\cite{hauert:2003,santos:2008} generalized it to other lattices as
well as to heterogeneous scale-free (SF) \cite{boccaletti:2006}
networks.

Within the above context, and following the seminal work by Santos
{\em et al.}  \cite{santos:2008} about the PGG on top of complex
networks, it is widely accepted that SF topologies enhance
considerably the emergence of cooperation as it was previously
observed for the Prisoner's Dilemma game
\cite{santos:2005,poncela:2008b}. In the case of the PGG
\cite{santos:2008} it is assumed that each of the groups in which the
game is played is automatically defined by considering each player and
all of her neighbors in the network. Following this recipe, several
mechanisms aimed at further enhancing cooperation on SF networks have
been investigated
\cite{Rong:2009,Lei:2010,Cao:2010,Shi:2010,Liu:2010,Peng:2010,Yang:2009,Zhang:2010}. However,
the structure of most real networks \cite{Ahn:2010} reveals that the
above assumption about the structure of the interaction groups does
not hold. On the contrary, most real interaction networks comprise
many small modules of densely interconnected nodes in which group
interactions take place. More importantly, these small modules tend to
overlap only slightly \cite{Palla:2005}, so that a given node rarely
involves all its social acquaintances when collaborating in one
groupal task. Therefore, the existing works on the PGG on networks
have neglected the role of the mesoscale patterns, i.\ e., the groups
in which the PGG is played.

\begin{figure}
\begin{center}
\epsfig{file=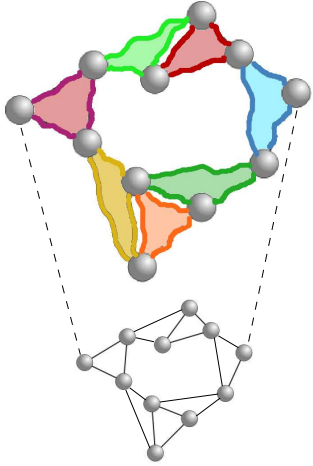,width=0.47\columnwidth,angle=0,clip=1}
\end{center}
\caption{A model network composed of $11$ nodes is shown. Each node
  engages in different groups of size $m=3$. The corresponding
  interaction backbone results appears projected below.}
\label{fig0}
\end{figure}

The aim of this Letter is to use the PGG to gain knowledge on the
effects of social heterogeneity (that arising when looking at the
number of social contacts of nodes) when the mesoscale structure of
small interaction modules is incorporated. Recent work by us
\cite{gomezgardenes:2011} hints that the fostering of cooperation
observed on heterogeneous networks may be different depending on
whether the mesoscopic structure of real collaboration networks is
incorporated or not. Thus, the question arises naturally as to what
are the effects of social heterogeneity when a mesoscale structure is
incorporated in synthetic models of homogeneous (Erd\"os-R\'enyi
\cite{boccaletti:2006}) and heterogeneous SF networks. Thus, the
answer to this question allows to disentangle the influence of social
heterogeneity from that imposed to the group structure in
\cite{santos:2008}. Surprisingly, our results show that cooperation is
not an increasing function of social heterogeneity. On the contrary,
an homogeneous social structure may lead to larger levels of
cooperation depending on the game setting and the updating rule at
work. We believe that this is an important contribution to both
evolutionary games on graphs and dynamics involving group
interactions on networks, in so far as data on social collaborations
very often reveals a rather homogenous group structure embedded in
heterogeneous social networks.

\section{Interaction networks and group structure}

Let us start by introducing the complex networks in which the
evolutionary dynamics of the PGG will be implemented. In
Fig.~\ref{fig0} we show a model network composed of $11$ individuals
with a complex interaction backbone. The complex interaction backbone
described by the connections among pairs of nodes appears as the
projection of the seven interaction groups highlighted in
Fig.~\ref{fig0}. Each of these groups comprises three individuals and
represents the interaction groups in which collaborative tasks take
place. It is evident that the set of groups is enough to define
univocally the resulting complex backbone of interactions. However,
the information provided solely by the projected network does not
allow to recover the groups in which each individual has been involved
to produce the final topology.

The above example highlights that any dynamical process involving
group interactions, such as the PGG, cannot be treated from the
macroscopic interaction backbone but it demands to incorporate the
mesoscopic patterns arising from the integration of all the
interaction groups. To this aim, it is useful to represent the system
as a bipartite network \cite{gomezgardenes:2008} in which two types of
nodes coexist: individuals and groups. The bipartite nature of a
system such as the one shown in Fig.~\ref{fig0} is characterized by
two distributions: {\em (i)} the probability that an individual
participates in $q$ groups, $P(q)$, and {\em (ii)} the probability that
$m$ individuals take part of one group, $P(m)$. In order to construct
such bipartite graphs we will consider a method of network generation
inspired in a model \cite{ramasco:2004} aimed at mimicking the
structure of scientific collaboration networks. Inspired in this
context, where the interaction groups represent co-authored articles,
and also in real collaboration data
\cite{newman:2001a,newman:2001b,newman:2004}, we consider the size $m$
of the groups to be small and homogeneous (note that $m=3$ in
Fig.~\ref{fig0}).

As in \cite{ramasco:2004}, the construction of our networks relies on
an iterative process in which interaction groups are created
sequentially starting from an initial core of $m$ individuals (that by
itself constitutes the first group of the system) and a set of $(N-m)$
unconnected individuals. At each step of the process, a new individual
from the unconnected set defines a new group of size $m$ by choosing
its $(m-1)$ partners among the remaining $(N-1)$ individuals in the
system. In order to generate a family of networks interpolating
between homogeneous and heterogeneous topologies we adopt a similar
strategy to that of the model introduced in
\cite{gomezgardenes:2006}. For each of its $(m-1)$ choices, the
newcomer assigns a probability $\Pi_{i}$ to the other $(N-1)$
nodes. With probability $\alpha$ the newcomer makes the choice
completely random so that the probability $\Pi_i$ that a node $i$ is
selected by the newcomer is $\Pi_{i}=1/(N-1)$. On the other hand, with
probability $(1-\alpha)$, the newcomer selects a partner $i$
proportionally to the number of groups $i$ belongs to, $q_i$, so that
$\Pi_i=q_{i}/\sum_{j} q_{j}$. When the newcomer has selected $(m-1)$
partners, the new group is constituted. This process is iterated
$(N-m)$ times so that the final network is composed of $(N-m+1)$
groups and $N$ individuals. In our case we will fix $N=5000$.

\begin{figure}
\begin{center}
\epsfig{file=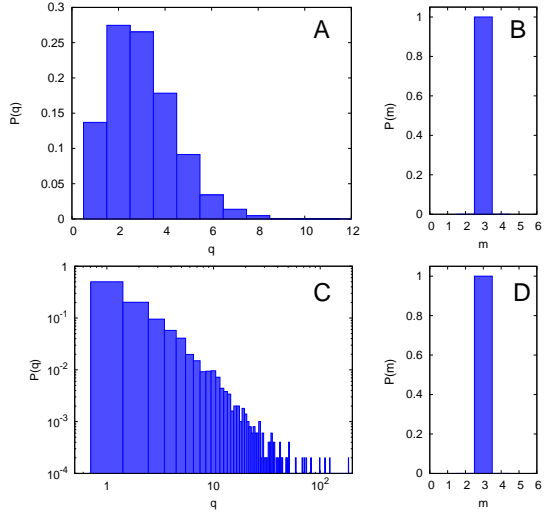,width=0.85\columnwidth,angle=0,clip=1}
\end{center}
\caption{Figures {\bf (A)} and {\bf (B)} show respectively the
  probability distribution for the number of groups each individual
  belongs to, $P(q)$, and the probability that a group is composed of
  $m$ players, $P(m)$, for those networks generated via random
  selection of group partners ($\alpha=1$). In {\bf (C)} and {\bf (D)}
  we show the two latter distributions for networks generated with
  preferential selection of group partners ($\alpha=0$).}
\label{fig1}
\end{figure}

Obviously, when $\alpha=1$ the network groups are always formed
following a random selection rule so that the final probabilty that a
node participates in $q$ groups follows a Poisson distribution
centered around $m$ (see Fig.~\ref{fig1}.A for which $m=3$) similarly
to Erd\"os-R\'enyi (ER) graphs. Alternatively, when $\alpha=0$ the
group structure is created by means of a preferential choice so that
the probability $P(q)$ follows a scale-free (SF) distribution,
$P(q)\sim q^{-3}$ with mean $\langle q\rangle=m$ (see
Fig.~\ref{fig1}.C). Let us recall that these two topologies are
homogeneous (ER) and heterogeneous (SF) only from the point of view of
the number of groups each individual participates in, {\em i.e.},
regarding the social structure. However in both topologies the group
structure is homogeneous, so that $P(m)$
is a delta function centered around $m$ (see Fig.~\ref{fig1}.B and
\ref{fig1}.D). Let us note that the projected versions of the
bipartite networks constructed with $\alpha=0$ and $\alpha=1$
correspond to SF and ER networks respectively. In the remainder of the
Letter we will focus on these two limiting cases (ER and SF) in order
to unveil the effects that social heterogeneity [as described by
  $P(q)$] has on the evolution of cooperation.

Having set the network structure we encode it by means of a
biadjacency matrix $\{B_{ji}\}$ [with $j=1,...,(N-m+1)$ and
  $i=1,...,N$] so that $B_{ji}=1$ if agent $i$ participates in group
$j$ and $B_{ji}=0$ otherwise. With this topological information one
can define the dynamics of the PGG as follows. At each time step $t$,
each individual $i$ plays a round of the PGG within each of the
$q_{i}$ groups it is engaged to. Obviously, the benefit obtained in
each of these games depends on the strategies of the $m$ agents
participating in each group. If we denote by $x_{i}^t$ the strategy of
agent $i$ during round $t$ of the PGG, so that $x_{i}^t=1$ when $i$
plays as cooperator and $x_{i}^t=0$ when $i$ defects, the overall
benefit after playing round $t$ of the PGG reads:
\begin{equation}
f_{i}(t)=\sum_{j=1}^{N-m+1}\frac{rB_{ji}}{m}\left[\sum^{N}_{l=1}B_{jl}x_{l}^tc_l \right]-x_{i}^tc_{i}q_{i}\;,
\label{bippayoff}
\end{equation}
where $q_{i}$ is (as defined above) the number of groups in which $i$
is engaged and $c_{i}$ accounts for the cost payed by agent $i$ in
each of her $q_{i}$ groups when playing as cooperator. We will study
two formulations of the PGG (as defined in \cite{santos:2008}). First,
we consider a fixed cost per game (FCG) formulation so that each
cooperator $i$ invests a fixed cost $c_i=z$ in each of the $q_i$
groups she participates in. Alternatively, we will also study the
situation of fixed cost per individual (FCI). In this latter scenario,
a cooperator invests a total amount $z$ that is equally distributed
among all her $q_i$ groups so that $c_i=z/q_i$

\begin{figure*}
\begin{center}
\epsfig{file=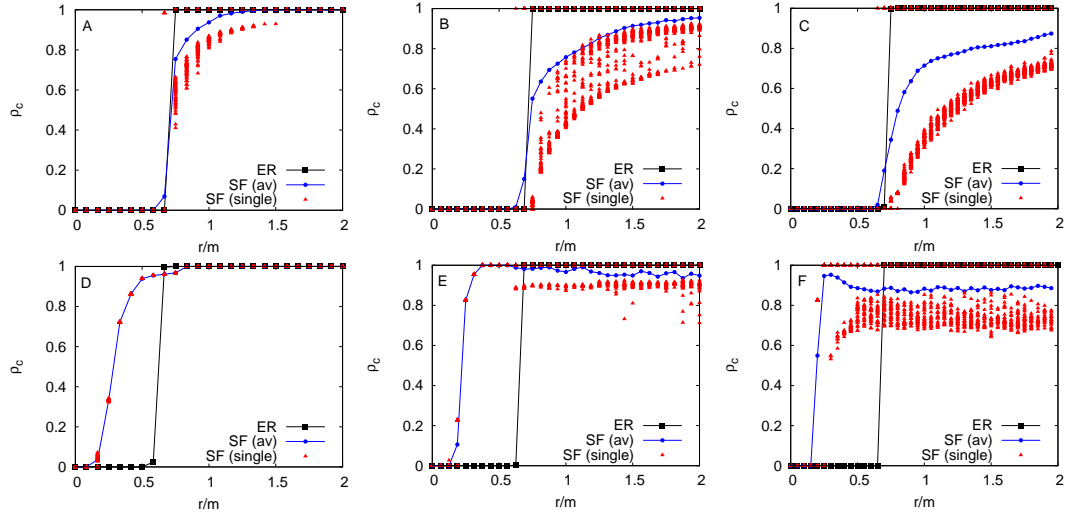,width=1.65\columnwidth,angle=0,clip=1}
\end{center}
\caption{Average fraction of cooperators, $\rho_c$, as a function of
  the normalized enhancement factor $r/m$ (we have fixed
  $z=1$). Panels A, B and C (top) are for the FCG formulation of the
  PGG while D, E and F (bottom) are for its FCI version. The panels
  correspond to $m=3$ (A and D), $m=4$ (B and E) and $m=5$ (C and
  F). The curves with filled squares correspond to ER networks while
  the results for SF one are those curved with filled circles. The
  triangles account for the level of cooperators reached in each of
  the $10^2$ realizations performed on SF networks for each value of
  $r/m$.}
\label{fig2}
\end{figure*}

Once a round of the PGG is played, every agent updates her strategy.
We will focus now in the replicator update rule as used in
\cite{santos:2008}. In this framework each agent $i$ chooses randomly
one of her partners, say $j$, and compares their benefits in the last
round of the game. If $f_{i}(t)\geq f_{j}(t)$ nothing happens and $i$
stays the same in the next round, $x_{i}^{t+1}=x_{i}^t$. However, when
$f_{j}(t)> f_{i}(t)$ agent $i$ will take the strategy of $j$
($x_{i}^{t+1}=x_{j}^t$) with probability:
\begin{equation}
P_{i\rightarrow  j}=\frac{f_{j}(t)-f_{i}(t)}{M}\;,
\end{equation} 
where $M$ is a normalization term that accounts for the difference
between the maximum possible payoff of $i$ and $j$ and the minimum
one. Thus, to compute $M$ one must calculate first the maximum and the
minimum payoff each individual can obtain, and use these values,
$f_{i}^{max}$ and $f_{i}^{min}$, to compute the correct $M$ associated
to each couple of nodes. In the case of the PGG with FCG we can take
advantage of the fixed size $m$ of the groups to derive analitically
the value of $M$ as a function of the number of groups each of the two
nodes, say $i$ and $j$, belongs to. In particular, when the $r\leq m$
we obtain:
\begin{equation}
M=\frac{z}{m}\max{[q_{i},q_{j}]}\left[m(r+1)-2r\right]\;,
\end{equation}
while for $r>m$ the expression for $M$ reads:
\begin{equation}
M=\frac{z}{m}\left[\max{[q_{i},q_{j}]}(m-1)r-\min{[q_{i},q_{j}]}(r-m)\right]
\end{equation} 
For the case of FCI, it is not possible to derive $M$ as a function of
$q_{i}$ and $q_{j}$ and one must compute the maximum payoff for each
node. This maximum payoff reads:
\begin{equation}
f_{i}^{max}=\frac{r}{m}\sum_{j=1}^{N-m+1} B_{ji}\sum_{l}B_{jl}\frac{zx_{l}}{q_{l}}\;.
\end{equation}
On the contrary, the minimum possible payoff of a node in the FCI
formulation does not the depends on the properties of the node:
$f_{i}^{min}=z(r-m)/m$. Finally, let us note that the group structure
plays no role in this update stage, as it makes use of the (projected)
network of contacts.

\section{Homogeneous {\em vs} Heterogenous networks} 

We now focus on the evolution of the cooperation for networks with
homogeneous group structure and either SF or ER social patterns. To
this end we simulate the evolutionary dynamics of the PGG from an
initial condition in which the number of defectors and cooperators is
roughly the same and they are randomly distributed across the
network. For each value of the normalized enhancement factor $r/m$, we
iterate a large number of rounds of the PGG (typically $5\cdot10^4$)
and we measure the average fraction of cooperators, $\rho_c$, over a
time window of $10^4$ additional rounds. The results reported for each
value of $r/m$ are the average over $10^2$ different initial
conditions.

In the top panels of Fig.~\ref{fig2} we show the evolution of $\rho_c$
as a function of $r/m$ for the PGG in its FCG version. Each of the
plots corresponds to a different value of $m$, namely $m=3$ in
Fig.~\ref{fig2}.A, $m=4$ in Fig.~\ref{fig2}.B and $m=5$ in
Fig.~\ref{fig2}.C, and all them show the curves $\rho_c(r/m)$ for both
ER (filled squares) and SF (filled circles) topologies. The main
finding is that the average level of cooperation achieved on ER
substrates is remarkably larger than those observed on SF
architecctures. Specifically, while the onset of cooperation occurs
around the same value $r_c/m\simeq 0.5$ (regardless of the value of
$m$) for both ER and SF substrates, the sharp boost in the cooperation
of ER networks is in contrast of the slow increment observed for SF
networks, particularly for $m=4$ and $m=5$. This striking result
points out that the ability of SF to outperform the promotion of
cooperators on ER networks reported in \cite{santos:2008} is
intrinsically due to the entanglement of social and group
heterogeneities (in \cite{santos:2008} the associated distribution of
group sizes in SF networks is $P(m)\sim m^{-\gamma}$, $\gamma$ being
the same exponent of the degree distribution of the SF network of
contacts). In our setting, the discrimination of social and group
heterogeneities in SF networks and the addition of a realistic group
architecture leads to a dramatic change in the ability of
heterogeneous networks to foster cooperation.

The differences in the average level of cooperation are not the unique
difference between homogeneous and heterogeneous networks. In the
panels of Fig.~\ref{fig2} we show the values reached by $\rho_c$ in
each of the realizations for SF networks. It is clear that, after the cooperation
onset, $r_c/m$, some of the realizations reach full cooperation while
others end up in a dynamical equilibrium in which cooperators and
defectors coexist. Note also that the value of $\rho_c$ associated to
those solutions displaying coexistence decreases significantly with
$m$. On the other hand, ER networks always lead to fixation, {\em
  i.e.}, the dynamics always reaches one absorbing state (either full
defection or full cooperation).

The results obtained with the FCI formulation are shown in the bottom
panels of Fig.~\ref{fig2}. This scenario turns out to favor the
emergence of cooperation on SF networks since its onset anticipates
significantly with respect to the FCG formulation (note that
$r_c/m\sim 0.2$ for all the values of $m$). On the contrary, for ER
the onset of cooperation takes place at the same value of $r/m$ as in
the FCG formulation. The enhancement shown by SF networks is clearly
due to the fact that in the FCI setting cooperators pay the same cost
regardless the number of groups they belong. This equivalence among
degree-classes allows cooperator hubs to collect more payoff while,
for defector hubs, the change from FCG to FCI does not represent any
improvement. On the other hand, the dynamical differences between SF
networks and ER networks persist since SF allows coexistence of
cooperators and defectors while ER do not. Moreover, as $m$ grows the
frequency of the solution displaying coexistence increases and for
large values of $r/m$ the average value of $\rho_c$ is lower than that
reached by ER networks for $m=4$ and $m=5$. Therefore, in the FCI
setting, when both the size of the groups $m$ and the degree of
enhancement $r$ increase, ER substrates outperform the ability of SF
networks to sustain cooperation.

\begin{figure}[t!]
\begin{center}
\epsfig{file=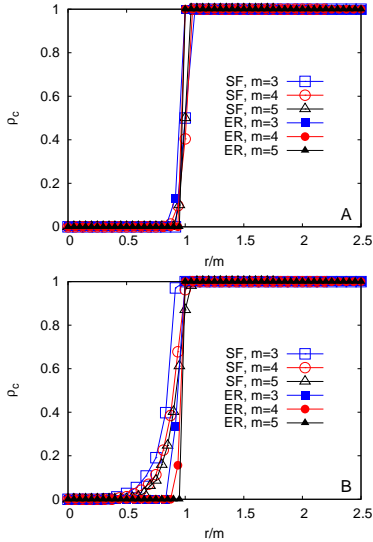,width=0.60\columnwidth,angle=0,clip=1}
\end{center}
\caption{We show the curves $\rho_c(r/m)$ for the PGG (we have fixed
  $z=1$) with Moran selection on ER (filled symbols) and SF (bold
  symbols) networks. Different group sizes are shown: $m=3$ (squares),
  $m=4$ (circles) and $m=5$ (triangles). Panel A (top) is for the PGG
  in the FCG setting while B (bottom) accounts for its FCI version.}
\label{fig4}
\end{figure}

\section{Other update rules}
To complete our study, let us now analyze the PGG
with two other update rules, namely, Moran selection and Unconditional
Imitation (UI). In the first case, a Moran agent $i$ chooses one
neighbor $j$ proportionally to her payoff (not randomly as in the
Replicator case). Then, agent $i$ copies the strategy of agent $j$ for
the next round, $x_{i}^{t+1}=x_{j}^t$, even if $j$ has performed
worse than $i$ [$f_{i}(t)>f_{j}(t)$]. Therefore, the probability that
$i$ takes the strategy of $j$ reads:
\begin{equation}
P_{i\rightarrow j}=\frac{f_{j}(t)}{\sum_{\langle j,i\rangle}f_{j}(t)}\;,
\end{equation}
where the symbol $\langle j,i\rangle$ means that the sum is over the
partners of $i$. Note that, at variance with the Replicator rule,
Moran selection allows mistakes. In the setting of UI the strategy
update is done as follows. After every round each agent imitates the
neighbour with the largest payoff, provided it is larger than her
own. Thus, at variance with Moran selection and the Replicator update,
UI is a completely deterministic rule while UI (as the Replicator
rule) does not allow mistakes. Note also that both Moran and UI are
context-focused rules (agents look at all their partners), while the
Replicator update is link-focused (agents look one randomly chosen
partner).

\begin{figure}
\begin{center}
\epsfig{file=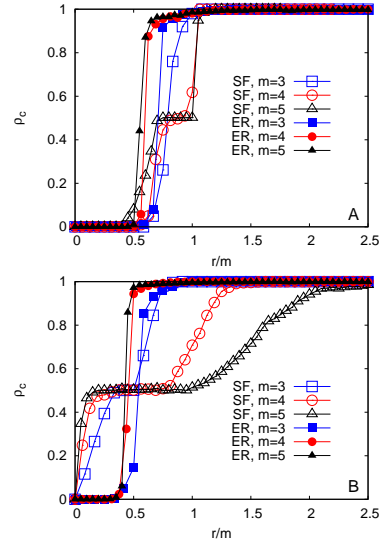,width=0.60\columnwidth,angle=0,clip=1}
\end{center}
\caption{We show the curves $\rho_c(r/m)$ for the PGG (we have fixed
  $z=1$) with the UI rule on ER (filled symbols) and SF (bold symbols)
  networks. Different group sizes are shown: $m=3$ (squares), $m=4$
  (circles) and $m=5$ (triangles). Panel A (top) is for the PGG in the
  FCG setting while B (bottom) accounts for its FCI version.}
\label{fig5}
\end{figure}

In Fig.~\ref{fig4} the results obtained with the Moran update are
shown. It is worth noticing that in FCG version (top panel), there is
no difference at all in behavior between SF and ER, while in FCI
(bottom panel) the onset of cooperation appears earlier in
heterogeneous SF networks, as with the Replicator rule, but in here
the differences are not as pronounced as in the former
case. Therefore, under Moran selection the degree of heterogeneity of
social interactions plays little role in the promotion of
cooperation. In Fig.~\ref{fig5} we show the behavior of systems with
UI update rule.  For the FCG situation (top panel) the cooperation
onset of SF and ER occurs simultaneously (around $r_c/m\simeq
0.5$). However, ER reaches full cooperation faster. In this sense, for
this cost scheme and under UI, ER promotes cooperation better than SF
(as in the Replicator case). On the other hand, for the PGG with FCI
(bottom panel) cooperation on SF networks rises suddenly from $r=0$
(even earlier than in the Replicator case); however, the ER network
reaches full cooperation much before than SF, particularly for $m=4$
and $m=5$. Interestingly, a plateau around $\rho_c\simeq 0.5$ appears
in both FCG and FCI cases for the SF network. For such values of $r/m$
where the plateau is observed the evolutionary dynamics ends up
suddenly (few generations after the initial condition) reaching either
a full cooperation or a full defection state. Thus, the final outcome
depends strongly on the initial configuration of strategies so that
for a large number of realizations $\rho_c\rightarrow\rho_c(t=0)=0.5$.
We believe that this is connected to the initial distribution of
strategies in the hubs as under this update scheme their behavior is
very determinant for the rest of the network.

\section{Conclusions}
The work reported here allows us to draw two important conclusions.
First, the enhancement of cooperation observed in PGG on SF networks
with respect to ER networks does not appear when taking into account
the details of the group structure of the population. This is clearly
so in the FCG scheme, while under FCI we observe that cooperation sets
on quite earlier but as the group size increases it becomes more
difficult to reach full cooperation.  Second, SF and ER networks
behave differently depending on the evolutionary dynamics under
consideration. Thus, the above comments apply to Replicator dynamics,
but Moran selection gives rise to basically similar behavior on both
types of networks and UI reverses the Replicator dynamics outcome,
with SF networks performing worse in general than ER
networks. Therefore, as has been shown for other social dilemmas
\cite{roca:2009a}, the outcome of a PGG on a network is far from
universal and depends on the network structure, on the evolutionary
dynamics and on the (mesoscale) group structure, a novel factor
arising from the game itself. We believe that this conclusion has far
reaching implications, the most important one being the relevance of
the network hierarchical structure for the emergence of cooperation in
a multi-player setup. Indeed, our results strongly indicate that when
trying to model cooperative behavior, the outcome of the model may
depend very much on this mesoscale structure which, in turn, are
almost always determined by the kind of function or cooperative
enterprise the agents are involved in.  In this context, it becomes
apparent that disentangling the scales associated to the different
types of relationships between the agents is crucial in order to
understand the observations in different social contexts.

\acknowledgments J.\ G.-G. is supported by the MICINN through the
Ram\'on y Cajal program and grants FIS2008-01240 and MTM2009-13838.
D.\ V.\ is supported in part by a postdoctoral contract from
Universidad Carlos III de Madrid.  A.\ S.\ was supported in part by
grants MOSAICO and Complexity-NET RESINEE (Ministerio de Ciencia e
Innovaci\'on, Spain) and MODELICO-CM (Comunidad de Madrid, Spain).


\end{document}